%% file: main.tex
\documentclass[]{eptcs}
\providecommand{\event}{Graphs as Models (GaM) 2016} % Name of the event you are submitting to
\usepackage{graphicx}

\usepackage{amsmath}
\usepackage{amssymb}
\usepackage{amsthm}
\usepackage{thmtools}

\newtheorem{definition}{Definition}
\newtheorem{example}{Example}

\newenvironment{mydescription}[2]{
\begin{list}{}{%
        \settowidth{\labelwidth}{#2{#1}}  % 
        }}%
{\end{list}} %
\newcommand{\MDB} {\begin{mydescription}}
\newcommand{\MDE} {\end{mydescription}}

\usepackage{todonotes}

\newcommand{\type}[1]{\textsl{#1}}
\newcommand{\elmt}[1]{\textsl{#1}}
\newcommand{\feature}[1]{\textsl{#1}}

\usepackage{listings}

\title{An EMOF-Compliant Abstract Syntax for Bigraphs}

\def\titlerunning{An EMOF-Compliant Abstract Syntax for Bigraphs}
\def\authorrunning{Timo Kehrer, Christos Tsigkanos \& Carlo Ghezzi}

\author{Timo Kehrer, Christos Tsigkanos and Carlo Ghezzi
  \institute{Dipartimento di Elettronica, Informazione e Bioingegneria\\Politecnico di Milano, Italy}
}

\begin{document}
\maketitle

%%%%%%%%%%%%%%%%%%%%%%%%%%%%%%%%%%%%%%%%%%%%%%%%%%%%%%%%%%%%%%%%%%%%%%%%%%%%%%%
%% Abstract
%%%%%%%%%%%%%%%%%%%%%%%%%%%%%%%%%%%%%%%%%%%%%%%%%%%%%%%%%%%%%%%%%%%%%%%%%%%%%%%

\begin{abstract}
Bigraphs are an emerging modeling formalism for structures in ubiquitous computing. 
 Besides an algebraic notation, which can be adopted to provide an algebraic syntax
for bigraphs, the bigraphical theory introduces a visual concrete syntax which 
is intuitive and unambiguous at the same time; the standard visual notation can be customized and thus tailored to domain-specific requirements.
 However, in contrast to modeling standards based on the Meta-Object Facility (MOF)
and domain-specific languages typically used in model-driven engineering (MDE), 
the bigraphical theory lacks a precise definition of an abstract syntax for 
bigraphical modeling languages.
 As a consequence, available modeling and analysis tools use proprietary 
formats for representing bigraphs internally and persistently, which hampers the 
exchange of models across tool boundaries.
 Moreover, tools can be hardly integrated 
with standard MDE technologies in order to build sophisticated tool chains 
and modeling environments, as required for systematic engineering of large systems
or fostering experimental work to evaluate the bigraphical theory in real-world applications.
 To overcome this situation, we propose an
abstract syntax for bigraphs which is compliant to the Essential MOF (EMOF)
standard defined by the Object Management Group (OMG).
 We use typed graphs as a formal underpinning of EMOF-based models and present a 
canonical mapping which maps bigraphs to typed graphs in a natural way.
 We also discuss application-specific variation points in the graph-based 
representation of bigraphs. 
 Following standard techniques from software product line engineering,
we  present a framework to customize the graph-based representation
to support a variety of application scenarios.
% We showcase our approach through
%  We implemented the approach as an extension of the popular Eclipse IDE 
% integrating the bigraph modeling tool Big Red  
% with the Eclipse Modeling Framework, 
% the de-facto reference implementation of OMG's EMOF standard.

% The approach is illustrated using a standard example of a context-aware 
% printer system. 
\end{abstract}

%%%%%%%%%%%%%%%%%%%%%%%%%%%%%%%%%%%%%%%%%%%%%%%%%%%%%%%%%%%%%%%%%%%%%%%%%%%%%%%
%% Paper
%%%%%%%%%%%%%%%%%%%%%%%%%%%%%%%%%%%%%%%%%%%%%%%%%%%%%%%%%%%%%%%%%%%%%%%%%%%%%%%

\input{intro.tex}

\input{bigraphs.tex}

\input{graphs.tex}

\input{mapping.tex}

\input{variations.tex}

\input{tooling.tex}

\input{conclusion.tex}

\bibliographystyle{eptcs}
\bibliography{literature}
\end{document}

%% file: intro.tex
%!TEX root = ./paper.tex

\section{Introduction}
Bigraphs and bigraphical reactive systems  have been proposed by 
Milner~\cite{Milner.Bigraphs.2009} as a fundamental theory for structures in 
ubiquitous computing~\cite{weiser1991computer}.
 A concise model of space, which forms the 
context of a ubiquitous system, serves as a starting point in the bigraphical 
modeling methodology.
 The idea of bigraphs is based on two fundamental concepts of discrete spaces: 
locality and connectivity, called \emph{placing} and \emph{linking} in 
bigraphical terminology.
 In essence, a bigraph consists of a \emph{place graph}, a forest defined over a 
set of nodes which is intended to represent entities 
and their locality in terms of a containment structure, and a \emph{link graph}, 
a hypergraph composed over the same set of nodes representing arbitrary linking among 
those entities.
 Possible local reconfigurations of the space are expressed as a set of 
reaction rules yielding a Bigraphical Reactive System (BRS). A reaction rule 
is a rewriting rule which consists of two bigraphs, called \emph{redex} 
and \emph{reactum}. Roughly speaking, the redex specifies a bigraphical pattern 
whose occurrence found in a ``host bigraph'' enables 
replacement by the reactum.
 
The bigraphical theory has been shown to be general enough as a meta-calculus to 
embed and thus unify reasoning over a set of existing formalisms and calculi
targeting different facets of computation, such as Petri nets~\cite{leifer2006transition}, the Ambient Calculus~\cite{Cardelli98mobileambients} and the pi-calculus~\cite{bundgaard2006typed}.
Bigraphs have found application in ubiquitous systems in diverse domains such as cyber-physical~\cite{sescps,tsigkanos2015ariadne} and biological systems~\cite{krivine2008stochastic}. The use of bigraphs and BRS as underlying supporting formalisms enables a spectrum of reasoning approaches spanning from modeling to simulation and verification.
However, as argued by Milner~\cite{Milner.Bigraphs.2009}, much experimental 
work has to be done in order to evaluate whether the idea of bigraphs serves as a 
suitable foundation for building large-scale ubiquitous systems. In this regard,
the bigraphical modeling approach shall be not only used for the purpose of
analysis, but even as a language for simulation and 
``programming''.
 Following this line of research, the theory of bigraphs
actually serves as a foundation for domain-specific languages.
 Besides an algebraic notation, which can be adopted to provide a textual 
syntax, the bigraphical theory introduces a visual concrete syntax 
%which is intuitive and unambiguous at the same time. 
which can be customized to
domain-specific requirements.

% On the modeling aspect for example, Big Red has focused on the 

Several approaches utilizing bigraphs exist, focusing on different aspects. % such as model checking, simulation or execution. % and reflected in a plethora of tools.
Visual design of bigraphical models has been considered~\cite{faithfull2013big}, while facilities for simulation, verification and execution are provided by a plethora of approaches utilizing  different forms of bigraphs and their dynamics. 
For example, %BPLTool~\cite{hojsgaard2011bpl} enables reasoning on BRS systems with binding, and has an emphasis on implementation correctness; 
DBtk\cite{bacci2009dbtk} targets a form of bigraphs with directed edges; SBAM~\cite{hojsgaard2012bigraphical} extends bigraphs with stochastic semantics; finally, BigMC~\cite{perrone2012model} performs model checking of matching properties.

%  target distributed BRS simulations.
% However, tool support is still in an early prototype stage.
% Available research prototypes focus on different aspects such as model 
% checking~\cite{perrone2012model},\cite{sevegnani2015bigraphs}, simulation or execution~\cite{mansutti2014towards}
%  visual editing \cite{faithfull2013big}, but are yet poorely integrated. 

Systematic engineering of systems using bigraphs, however, is hindered by obstacles to modeling, integration, evaluation, and the overall development process.
Bigraphical tool support is still in its infancy and available research prototypes are yet poorly integrated. Methods supporting development of large-scale models are sorely missing. 
% Systematic engineering of systems is is hindered.
% This is a major obstacle 
% in 
 % in actual 
This situation hampers empirical assessments with real users to evaluate expressive power, generality, and utility of the bigraphical theory in real-world applications.
 An important reason for this unfortunate situation in actual engineering of systems is that the notion of 
bigraphs as a domain-specific modeling language yet lacks a precise and
commonly accepted definition of an abstract syntax.
 The available modeling and analysis tools use proprietary formats for 
representing bigraphs internally and persistently, which hampers the exchange of 
models across tool boundaries and the development of sophisticated tool chains.

\paragraph{Contributions.}
% To overcome this unfortunate situation, t
In this paper, we propose an abstract syntax 
for bigraphs which is compliant with the Essential Meta-Object Facility (EMOF) standard defined by the Object Management Group (OMG)~\cite{MOF}.
This facilitates the integration of bigraphical modeling and analysis tools
with mainstream technologies for model-driven engineering (MDE), 
typically based on the EMOF standard.
 We use typed graphs as a formal underpinning of EMOF-based models and present 
a canonical mapping which maps bigraphs to typed graphs in a natural way.
 Thereupon, we discuss application-specific variation points in the graph-based 
representation of bigraphs. 
 Following techniques from software product line engineering,
we present a framework to customize the graph-based representation to support a variety of applications.
 We implemented the approach as an extension of the Eclipse IDE 
integrating a bigraph modeling tool~\cite{faithfull2013big} with the 
Eclipse Modeling Framework \cite{EMFBook}, the de-facto reference implementation 
of EMOF.%OMG's EMOF standard.

\paragraph{Paper structure.}
%The rest of the paper is structured as follows.
 Section \ref{sec:bigraphs} presents bigraphs, along with a motivating example of 
a cyber-physical space which is used throughout the paper.
 Section \ref{sec:graphs} briefly introduces our notion of typed graphs which 
is used in Section \ref{sec:mapping} for a canonical mapping of bigraphs to 
typed graphs. 
 Section \ref{sec:variations} bootstraps a set of application-specific 
variation points of a bigraphical abstract syntax, aimed to support different 
forms of bigraphs. 
 Section~\ref{sec:tooling} outlines how to integrate the approach into an
existing MDE environment and presents an example application.
 Section \ref{sec:conclusion} concludes the paper along with an outlook on
future work.

%% file: bigraphs.tex
%!TEX root = ./paper.tex

\section{The Idea of Bigraphs: Informal Introduction and Visual Syntax}
\label{sec:bigraphs}

A \emph{Bigraph} consists of two graphs. A \emph{place graph} is a
forest, a set of rooted trees defined over a set of nodes.
A \emph{link graph} is a hypergraph composed over the same set of nodes
and a set of edges, each linking an arbitrary number of nodes.
Connections of an edge with its nodes are called ports. Place and
link graphs are orthogonal, and edges between nodes can cross locality
boundaries. The types of nodes, called controls in bigraphical terminology,  are defined by a \emph{signature}.
To avoid misunderstandings, hereafter we use the terms and definitions introduced by Milner~\cite{Milner.Bigraphs.2009}, although we acknowledge that sometimes they may sound unconventional and unintuitive.

A bigraph $B$ is a structure $(V_B, E_B, ctrl_B, prnt_B, link_B) : \langle k,X \rangle \to \langle m,Y \rangle$ where $V_B$ 
is a set of nodes, $E_B$ is a set of edges, $ctrl_B$ is the control map that assigns controls to nodes, $prnt_B$ is 
the parent map that defines the nesting of nodes, and $link_B$ is the link map that defines a link structure.
The notation $\langle k,X \rangle \to \langle m,Y \rangle$ indicates that the bigraph has 
$k$ sites which denote the presence of other unspecified nodes, $m$ roots, a set $X$ of inner names and a set $Y$ of of outer names. 
$\langle k,X \rangle$ and $\langle m,Y \rangle$ are called the inner and outer
interfaces (or faces) of the bigraph.

% These are respectively known as the inner and outer interfaces of the bigraph.
The bigraphs presented here are pure (non binding) and concrete; placing and
linking are independent structures, and nodes and edges have discrete
identifiers~\cite{milner2006pure}. Moreover, signatures are basic signatures abstracting from dynamic aspects, namely activeness and passiveness of controls.
% In the following, we mean pure bigraphs and basic signatures when we speak about bigraphs and signatures, respectively.
We present only essential structural definitions; the interested reader can refer to~\cite{Milner.Bigraphs.2009}.
Names, node-identifiers and edge-identifiers are drawn from infinite sets,  respectively $\mathcal{X}$ , $\mathcal{V}$ and $\mathcal{E}$, disjoint from each other.

% (basic signature) 
\begin{definition}[Signature]
A signature $\Sigma$ is a pair ($\kappa$, ar). It consists of a
set $\kappa$ whose elements are node types called controls, and a function 
$ar : \kappa \rightarrow \mathbb{N}$ assigning a natural number representing arity (the number of ports) to each 
control.
\end{definition}

In the sequel, we will describe a signature as a set of pairs $control : arity$.

% (concrete place graph) 
\begin{definition}[Place Graph]
A place graph $B^P$ over a signature $\Sigma$ is a structure
$B^P = (V_B , ctrl_B , prnt_B ) : k \rightarrow m$
having an inner face $k$ and an outer face $m$, both finite ordinals 
$k = \{0, 1, \dots, k-1\}$ and $m = \{0, 1, \dots, m-1\}$; 
these index the sites and roots of the place graph. $B^P$ has a finite set $V_B \subset \mathcal{V}$ of nodes,
a control map $ctrl_B : V_B \rightarrow \kappa$ and a parent mapping
$prnt_B : k \cup V_B \rightarrow V_B \cup m$
which is acyclic.%, i.e. if prnt iV (v) = v for some v ∈ V_B then i = 0.
\end{definition}

\begin{definition}[Link Graph]
% (concrete link graph) 
A link graph $B^L$ over a signature $\Sigma$ is a structure
$B^L = (V_B , E_B , ctrl_B , link_B ) : X \rightarrow Y$
with finite inner face $X \subset \mathcal{X}$ and finite outer face $Y \subset \mathcal{X}$, called the
names of the link graph. $B$ has finite sets $V_B \subset \mathcal{V}$ of
nodes and $E_B \subset \mathcal{E}$ of edges, a control mapping $ctrl_B : V_B \rightarrow \kappa$ and a link mapping
$link_B : X \cup P_B \rightarrow E_B \cup Y$
where $P_B \overset{def}{=} \{(v, i)~|~ i \in ar(ctrl_B (v))\}$ is the set of ports of B; thus $(v, i)$ is the $i^{th}$ port of node $v$.
%We call $X \cup P_B$ the points of $B^L$, and $E_B \cup Y$ its links.
\end{definition}

A bigraph is then a pair of a place graph and a link graph, called its \emph{constituents}.

\begin{definition}[Bigraph]
A bigraph 
$B = (V_B , E_B , ctrl_B , prnt_B , link_B ) : \langle k, X \rangle \rightarrow \langle m, Y \rangle$ 
over a signature $\Sigma$
consists of a place graph $B^P = (V_B , ctrl_B , prnt_B ) : k \rightarrow m$ and a
link graph $B^L = (V_B , E_B , ctrl_B , link_B ) : X \rightarrow Y$.
\end{definition}

Throughout the paper, we refer to the nodes, edges, roots, sites, inner names, outer names 
and ports as the \emph{elements} of a bigraph.
Moreover, we use the rigorous visual notation introduced in~\cite{Milner.Bigraphs.2009} as concrete syntax for bigraphs.
In this notation, roots and sites are shown as dotted and shaded boxes, respectively,
with indices defined by the place graph's outer and inner face
attached to them.
Nodes are represented as ellipses with a control label attached where their containment relation is visually represented by nesting a place graph node inside another one.
Ports of nodes are represented as black bullets; they can be linked together to form edges.
Moreover, ports can be linked to outer names, and edges can be linked to inner names of the link graph interface.
By convention, outer names are drawn above a bigraph figure while inner names are drawn below. 

In the following, we introduce a sample bigraph serving as running example. % throughout the paper.
% It is inspired by the context-aware printing system which is described in \cite{}.
Consider an instance of a cyber-physical space that models a printing scenario, inspired by~\cite{faithfull2013big}; an office environment comprised of two rooms, where one contains a computer and a printer, while the other one contains a user who holds a print job. A user can submit the job for printing through the computer connected to the printer.
% The model describes a building in which users can submit print jobs to a print spool, an then
% move into a room with any printer connected to that print spool, at which point the printer will
% complete the job. 
Example~\ref{examplebg} focuses on static aspects of this system, i.e.\ it shows
a snapshot at a particular point of time.

% #sites, {innernames}, #roots, {outernames}

\begin{figure}[ht]
 \centering
 \includegraphics[width=0.55\textwidth]{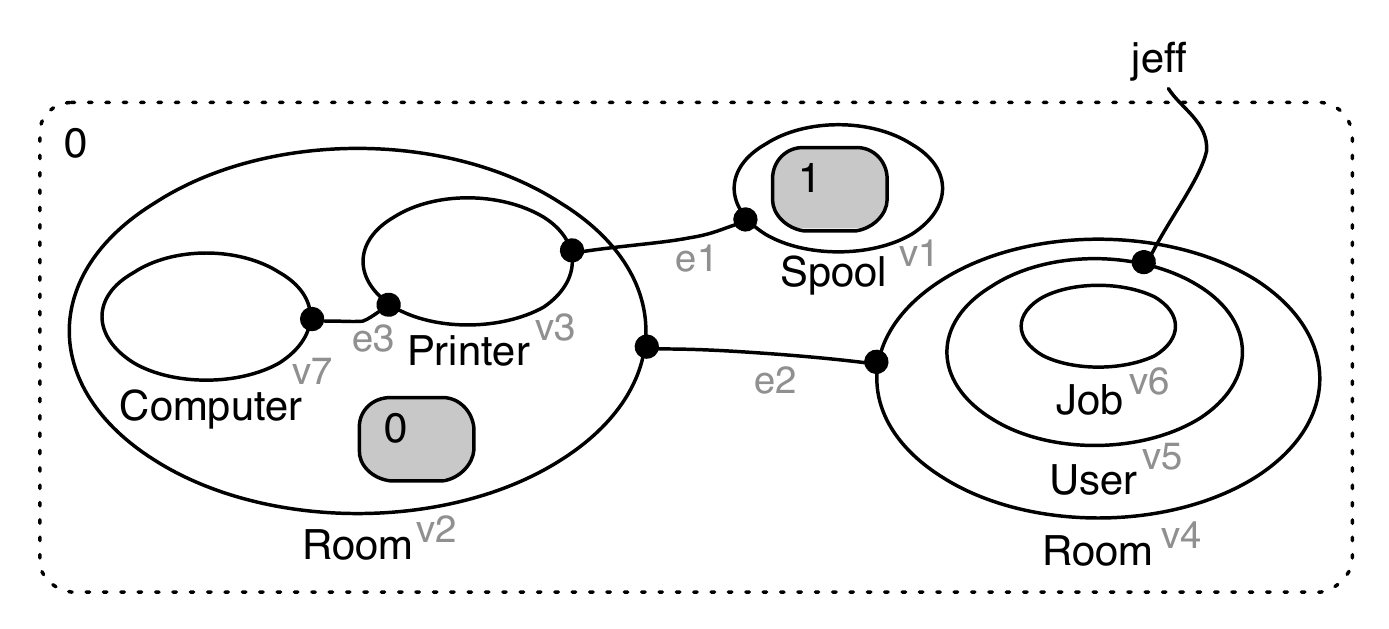}
 \caption{Bigraph model of a context-aware spooling system.}
 \label{fig:printer}
\end{figure}

% \todo[inline]{We should use node and edge identifiers, $v1, ...$ and $e1, ...$ in Figure \ref{fig:printer}, aligned with Figure \ref{fig:printer-graph}}

\begin{example}[Snapshot of a bigraph model of a context-aware printing system]
\label{examplebg}
Fig.~\ref{fig:printer} shows a visual representation of a bigraph 
$B_1 : \langle 2, \emptyset \rangle \rightarrow \langle 1, \{\text{jeff}\} \rangle$ over signature 
$\Sigma_1 = \{Job:0,\ User:1,\ Room:1,\ Spool:1,\ Printer:2,\ Computer:1\}$.
It consists of a single root and defines one outer name $jeff$.
The root contains nodes corresponding to a print $Spool$ and two $Room$s. 
The rooms are connected by an edge, the ports being linked to this
edge, semantically representing a connection that corresponds to room doors. The left $Room$ contains a $Printer$ which is connected  to the print $Spool$, by linking the ports of the $Printer$ and the print $Spool$ with a common 
edge. Moreover, a node of control $Computer$ is connected to the $Printer$. The $Room$ on the left and the $Spool$ contain sites, which serve as placeholders to denote the presence of unspecified nodes (representing unspecified entities).
% indicating that there might be more objects of the cyber-physical space in which we are currently not interested. 
The $Room$ on the right contains a $User$, in turn containing a print $Job$ which has not yet been 
submitted for printing. A link from the user's port to outer name jeff identifies the $User$. In Fig.~\ref{fig:printer}, grey labels identify nodes and edges for explanation purposes in following sections. 

\end{example}

%% file: graphs.tex
%!TEX root = ./paper.tex

\section{Graph-based Representation of EMOF Models}
\label{sec:graphs}
In MDE, the conceptual contents of a model are usually considered as a typed 
attributed graph which is known as the abstract syntax graph (ASG) of a model.
% It abstracts from conceptually irrelevant details, e.g., the diagram
% layout in the case of visual models.
 The allowed types of nodes and edges in an ASG as well as their possible 
relations are typically defined by a meta-model which thus serves as an abstract 
syntax specification. 
 In the remainder of this section, we introduce our graph-based representation 
of meta-models and instance models.
 The formalization is based on \cite{biermann2012formal,ehrig2009generating} and 
is fully compatible with the EMOF standard.

\paragraph{Graphs and graph morphisms.}
Following object-oriented modeling principles of the EMOF standard,
our basic notion of a graph refers to a directed unlabelled graph  
$G = (G_N, G_E, src_G, tgt_G)$ 
which consists of a set $G_N$ of nodes, a set $G_E$ of edges, as well as
source and target functions $src_G, tgt_G : G_E \rightarrow G_N$ 
associating to each edge its source and target node.
 Given two graphs $G$ and $H$, a pair of functions $(f_N, f_E)$ with 
$f_N : G_N \to H_N$ and $f_E : G_E \to H_E$, forms a graph morphism 
$f : G \to H$ if it preserves source and target, i.e.\ 
$f_N \circ src_G = src_H \circ f_E$ and 
$f_N \circ tgt_G = tgt_H \circ f_E$. In other words, for each edge $e_G \in G_E$ 
there is a corresponding edge $e_H = f_E(e_G) \in H_E$
such that $src_G(e_G)$ is mapped to $src_H(e_H)$ and $tgt_G(e_G)$ is mapped to $tgt_H(e_H)$.
% \begin{itemize}
%  \item $src_G(e_G)$ is mapped to $src_H(e_H)$, and 
%  \item $tgt_G(e_G)$ is mapped to $tgt_H(e_H)$.
% \end{itemize}
% A graph morphism is a monomorphism if $f_N$ and $f_E$ are injective functions. 
% A graph morphism is an isomorphism if $f_N$ and $f_E$ are bijective functions.

\paragraph{Meta-models as type graphs.}
A meta-model is formally treated as a distinguished graph called \emph{type graph}, 
whose nodes and edges represent \emph{node types} and \emph{edge types}, respectively.
 In addition, a type graph comprises the definition of an \emph{inheritance hierarchy}
including \emph{abstract node types}, a \emph{containment structure},  
\emph{opposite edges} representing bidirectional edge types,
and \emph{multiplicities} attached to edge types.

\begin{definition}[Type graph]\label{def:typegraph}
A type graph $TG = (T,I,A,C,OE,mult)$ is a graph 
$T = (T_N, T_E, src_T, tgt_T)$, representing node types and edge types,
equipped with the definition of a type hierarchy $I \subseteq T_N \times T_N$,
which must be an acyclic relation, a set $A \subseteq T_N$ identifying 
abstract node types, a containment structure $C \subseteq T_E$,  
and an anti-reflexive, symmetric and unique relation
{\em opposite edges} $OE \subseteq T_E \times T_E$ representing 
bidirectional edge types.
Moreover, multiplicities can be attached to edge types by function $mult : T_E \rightarrow Mult$. 
$Mult$ denotes the set of multiplicities, and a multiplicity is a 
pair $[lb,ub] \in \mathbb{N} \times (\mathbb{N} \cup \{*\})$ with $lb \leq ub$ or $ub = *$.  
\end{definition}
Given $(x,y) \in I$, $x$ refers to the {\em subtype} and $y$ to the {\em supertype} induced by 
the inheritance hierarchy.
We use function $allSub : T_N \rightarrow 2^{T_N}$ %from node types to their powerset, 
with 
$allSub(x) = \{y \in T_N \mid (y,x) \in I^+\}$ and $I^+$ being the transitive 
closure of the inheritance relation $I$, to refer to all subtypes of a node 
type.

\paragraph{Models as typed graphs.}
We formalize the typing relation between models and meta-models by a special 
graph morphism, called \emph{typing morphism}, relating a typed graph with its associated 
type graph.

\begin{definition}[Typed graph and typing morphism]\label{def:typed-graph}
An instance graph $G$ is typed over a fixed type graph $TG$ if there is a typing morphism
$type_G : G \rightarrow T$ which maps the nodes and edges of $G$ to those of $T$ 
in a suitable way, i.e.\ for each edge $e_G \in G_E$ there is a 
corresponding edge $e_{T} = type_G(e_G) \in T_E$, such that: 
\begin{itemize}
 \item $src_G(e_G)$ is mapped to $src_T(e_{T})$ or any subtype $x \in allSub(src_T(e_{T}))$, and 
 \item $tgt_G(e_G)$ is mapped to $tgt_T(e_{T})$ or any subtype $x \in allSub(tgt_T(e_{T}))$.
\end{itemize}
%The relation $type_G$ is called typing morphism.
\end{definition}

\vspace{0.25cm}
% The typing induces $G_C = \{e \mid type_G (e) \in C\}$ 
% and $contains_G  = \{(src_G(e), tgt_G(e)) \mid \forall e \in G_C\}\mbox{ }\cup\mbox{ } 
%                   \{(x, z) \mid \exists y \in G_N : (x,y) \in contains_G \wedge (y,z) \in contains_G)\}$.
% We say that graph $G$ typed over $TG$ is {\em valid}, if
% \begin{itemize}
% \item each node has {\em at most one container}: $\forall e_1,e_2 \in G_C : tgt_G(e_1) = tgt_G(e_2) \implies e_1 = e_2$, and
% \item has {\em no containment cycles}: $\forall x \in G_N : (x, x) \notin contains_G$, and
% \item {\em all opposite edges} are handled consistently: If $(e1,e2) \in OE$, then $\forall e \in G_E$
%           with $type_G(e) = e_1$ there is also an $e' \in G_E$ with
%           $type_G(e') = e_2$ and $src_G(e) = tgt_G(e')$ and $src_G(e') = tgt_G(e)$.
% \end{itemize}
% In the following, speaking about typed graphs, we assume them to be valid.

We say that graph $G$ typed over $TG$ is {\em valid} if there are
{\em no containment cycles}, each node has {\em at most one container},
and {\em all opposite edges} are handled consistently.
% , i.e.\ if $(e1,e2) \in OE$, then $\forall e \in G_E$
%           with $type_G(e) = e_1$ there is also an $e' \in G_E$ with
%           $type_G(e') = e_2 \mbox{ }\wedge\mbox{ } src_G(e) = tgt_G(e') \mbox{ }\wedge\mbox{ } src_G(e') = tgt_G(e)$.
In the sequel, speaking about typed graphs, we assume them to be valid.

\paragraph{Attributed graphs.} 
Throughout this paper, we largely neglect the formal treatment of node
attributes.
However, they can be handled by following the definition of attributed
graphs, which can be found in \cite{Heckel2002Attributed}. 
The main idea of formalizing node attributes is to consider them as edges of a 
special kind referring to \emph{(i)} data types in case of type graphs and 
\emph{(ii)} data values in case of typed graphs.

\paragraph{Notational conventions.}
We adopt the syntax of UML class and object diagrams as a compact visual notation 
for type and instance graphs, respectively.
 Visual connections between nodes without arrowheads denote bidirectional edges.
Although formally treated as edges of a special kind, attributes 
are shown as integral parts of nodes. 
 For node types, we use the notation $a:DT$ to denote the declaration of 
an attribute $a$ of type $DT$.
 For instance graph node attributes, we write $a=v$ to symbolize the assignment 
of value $v$ to attribute $a$. For a given node $x \in G_N$, we write $x.a$ to
refer to the value of its attribute $a$.

%% file: mapping.tex
%!TEX root = ./paper.tex

\section{A Canonical Mapping of Bigraphs to Typed Graphs}
\label{sec:mapping}

In this section, we present a canonical mapping of bigraphs to typed graphs.
We first introduce a basic type graph $TG_{Base}$ which models the constituents of 
bigraphs in a natural way, following standard object-oriented design principles.
 Subsequently, we show how to extend $TG_{Base}$ to a type graph $TG_\Sigma$
for a given bigraphical signature $\Sigma$.
 Having compatible type definitions $\Sigma$ and $TG_\Sigma$ at hand, we  
we can finally define the relation between a bigraph $B$ over $\Sigma$ and its 
corresponding graph $G$ typed over $TG_\Sigma$.
 
\subsection{Basic Type Graph}
\begin{figure}[ht]
 \centering
 \includegraphics[width=0.9\textwidth]{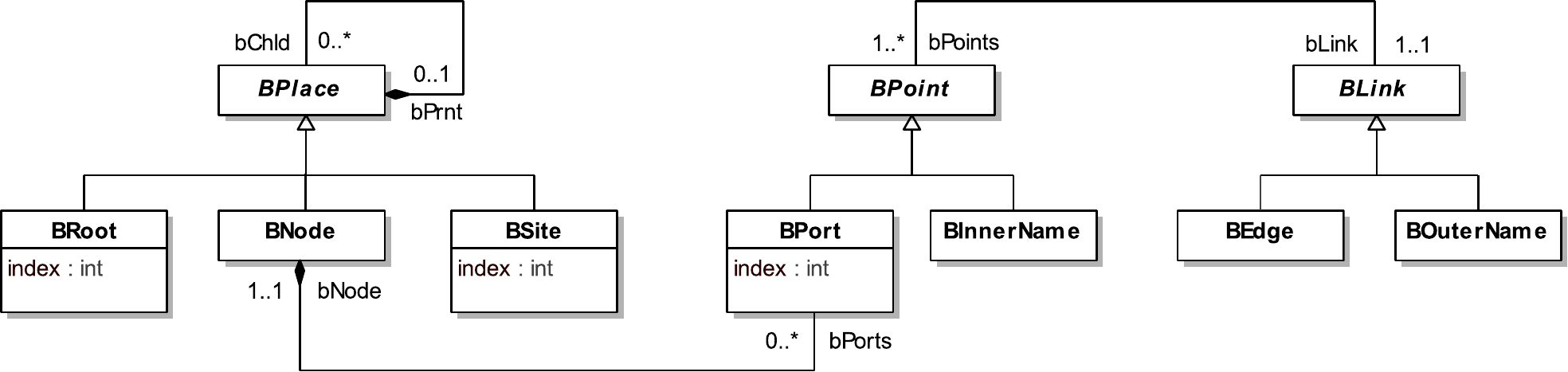}
 \caption{Basic type graph $TG_{Base}$ modeling the anatomy of bigraphs}
 \label{fig:TG-Base}
\end{figure}

A basic type graph modeling the core concepts of bigraphs %\footnote{The basic concepts are summarized in~\cite{Milner.Bigraphs.2009}--Fig.1 of Chapter 1--and called ``the anatomy of bigraphs''} 
is shown in 
Fig.~\ref{fig:TG-Base}.
% It defines node types for all kinds of bigraphical elements; edge types model
% the possible structural relations between them.
 The left-hand side of Fig.~\ref{fig:TG-Base} models the 
elements of the place graph. We have an abstract node type \type{BPlace} for 
representing places; the nesting of places is represented by instances of the  
edge type \type{bPrnt} and its opposite \type{bChld}. 
There are three concrete subtypes of \type{BPlace}, namely 
\type{BRoot}, \type{BNode} and \type{BSite} in order to represent the roots, 
nodes and sites of a bigraph, respectively. 
We use attributes \type{index:int} to index the roots and sites.
 The right-hand side of Fig.~\ref{fig:TG-Base} models the 
elements of the link graph. %We define the node types \type{BPoint} and \type{BLink}; 
An instance of \type{BPoint} is connected to exactly one instance of 
\type{BLink} via an edge of type \type{bLink}. Conversely, an instance of 
\type{BLink} is connected to one or more instances of type \type{BPoint} 
via edges of type \type{bPoints}. Edge types \type{bLink} and \type{bPoints}
are opposite to each other.
 Node types \type{BPoint} and \type{BLink} are abstract node types. Each of them
has two subtypes, namely \type{BPort} and \type{BInnerName} as well as \type{BEdge} and \type{BOuterName},
respectively. The containment edge type \type{bPorts} expresses that ports are contained
by nodes of type \type{BNode}; conversely, each node of type \type{BPort} has a parent 
node of type \type{BNode}, expressed by edge type \type{bNode}. Edge types 
\type{bPorts} and \type{bNode} are opposite to each other, thus defining a 
bidirectional edge type.
We use attribute \type{index:int} to index the ports in the graph-based representation
of a bigraphical node.
We will refer to the basic type graph as
$TG_{Base} = (T_{Base},I_{Base},A_{Base},C_{Base},OE_{Base}, mult_{Base})$ with
$T_{Base} = (T_{Base_N}, T_{Base_E}, src_{T_{Base}}, tgt_{T_{Base}})$.

% Node- and edge-identifiers are drawn from the finite set $\mathcal{T} = \{ 
% BPlace, BRoot, BSite, ...
% \}$
% 
% $TG_{Base} = (T_{Base},I_{Base},A_{Base},C_{Base},OE_{Base},mult_{Base})$
% with
% $T_{Base_N} = \{BPlace, BRoot, BNode, BSite \}$,
% $T_{Base_E} = \{ \}$, 
% $src_{T_{Base}} =  \{ \}$,
% $tgt_{T_{Base}} =  \{ \}$,
% $I_{Base} = \{ \}$,
% $A_{Base} = \{ \}$,
% $C_{Base} = \{ \}$,
% $OE_{Base} = \{ \}$,
% $mult_{Base} = \{ \}$.

\subsection{Mapping Bigraphical Signatures to Type Graphs}
The basic type graph introduced in Fig.~\ref{fig:TG-Base} is generic, i.e., it provides no information about bigraphical signatures: it neither models the controls nor the arities (number of ports) of nodes.
%the  defined by a signature nor  are not modeled by the basic
%type graph. 
% Secondly, a node representing a bigraphical node may contain an arbitrary number of 
%ports, i.e.\ arities defined by the signature are not considered.
%%
To overcome this, hereafter we extend the basic type graph 
for a given signature.
We handle controls through a
{\em control-compatible extension} of the basic type graph w.r.t.\ a given 
signature.
We then address arities via an additional well-formedness rule defined
over a control-compatible extension of the basic type graph.

\paragraph{Controls.}
Our control-compatible extension of the basic type graph w.r.t.\ a 
given signature introduces for each control defined by the signature 
a corresponding subtype of the generic node type 
\type{BNode}.

Let $\Sigma = (\kappa, ar)$ be a bigraphical signature and 
%$TG_\kappa = (T_\kappa,I_\kappa,A_\kappa,C_\kappa,mult_\kappa)$ with $T = (T{\kappa_N}, T{\kappa_E}, src_{T_\kappa}, tgt_{T_\kappa})$ be
$TG$ 
a type graph extending $TG_{Base}$, i.e.\ we have $TG \supseteq TG_{Base}$.
We say that $TG$ extends $TG_{Base}$ in a \emph{control-compatible} way 
w.r.t.\ $\Sigma$, if  
$T_N    = T_{Base_N} \cup \kappa$,
$T_E    =  T_{Base_E}$,
$src_T  =  src_{T_{Base}}$,
$tgt_T  =  tgt_{T_{Base}}$,
\mbox{$I      =  I_{Base} \cup \{(x,BNode) \mid x \in \kappa\}$},
$A      =  A_{Base}$,
$C      =  C_{Base}$,
$OE      =  OE_{Base}$, and
$mult   =  mult_{Base}$.
In the following, we write $TG_\kappa$ to emphasize the fact that a type graph
extends the basic type graph $TG_{Base}$ in a control-compatible 
way w.r.t.\ to a signature $\Sigma$.

Please note that bigraphical controls defined by $\Sigma$ and a subset 
of the identifiers of node types defined by $T_\kappa$ are drawn from the same set 
$\kappa$. To avoid confusion, we introduce an identity function $\Phi = Id_\kappa$ on set $\kappa$:
given a control $K \in \kappa$, we use the notation $\Phi(K)$ to refer to its
corresponding node type in $T_\kappa$.

\paragraph{Arities.}
To handle arities defined by a signature $\Sigma$, we introduce
an additional well-formedness rule defined over the control-compatible extension
of the basic type graph w.r.t.\ $\Sigma$.
For each node type $\Phi(K)$ we restrict the multiplicity $[0..*]$ 
defined by the generic edge type \type{bPorts} to a fixed value $ar(K)$.
%such that nodes of type $\Phi(K)$ must have exactly $ar(K)$ ports. 
 We use first order logic to formalize the respective well-formedness rule.

Let $\Sigma$ be a signature and $TG_\kappa$ be an extension of 
$TG_{Base}$, control-compatible w.r.t.\ $\Sigma$. 
For each graph $G$ typed over $TG_\Sigma$ the following condition must hold,
where $ran(\Phi)$ refers to the range of function $\Phi$:

\vspace{-0.5cm}
\begin{equation}\label{eq:wfmd-arities}
\forall n \in G_n : type_G(n) \in ran(\Phi) \implies \left\vert\{e \mid src_G(e) = n \wedge type_G(e) = bPorts\}\right\vert = ar(\Phi^{-1}(type_G(n)))
\end{equation}

\noindent
Given a node $G_n$ with $type_G(n) \in ran(\Phi)$,  
$\{e \mid src_G(e) = n \wedge type_G(e) = bPorts\}$
refers to its set of outgoing edges of type \type{bPorts}.
The cardinality of this set has to be equal to the 
aritiy of the corresponding control $\Phi^{-1}(type_G(n))$.

\paragraph{Bigraphical signatures as type graphs.}
Given a bigraphical signature $\Sigma$ and a type graph $TG$, we say that $TG$ is
compatible to $\Sigma$ if \emph{(i)} it extends the basic type graph $TG_{Base}$
in a control-compatible way and \emph{(ii)} control arities are properly 
represented as additional well-formedness rule
(\ref{eq:wfmd-arities}).
 In the following, we write $TG_\Sigma$ to indicate the fact that a type graph
is compatible w.r.t.\ to a signature $\Sigma$.

\begin{figure}[ht]
 \centering
 \includegraphics[width=1.0\textwidth]{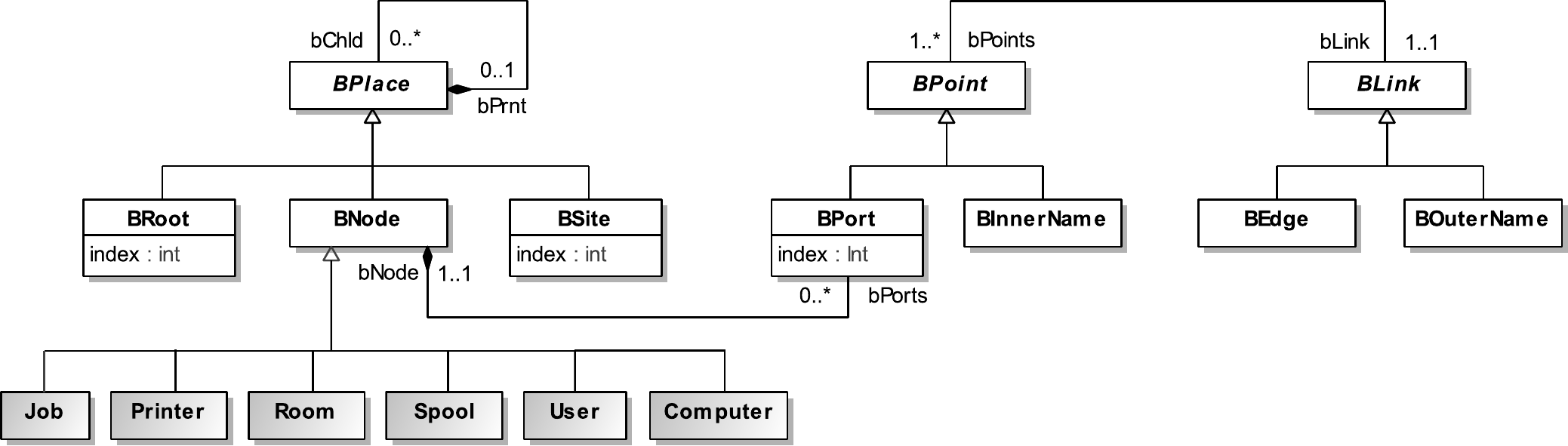}
 \caption{Type graph $TG_{\Sigma_1}$ compatible to signature $\Sigma_1$ of our running printer example}
 \label{fig:TG-Printer}
\end{figure}

\begin{example}[Bigraphical signature as type graph]
Fig.~\ref{fig:TG-Printer} shows how the bigraphical signature $\Sigma_1$ of our running printer example is represented
as type graph $TG_{\Sigma_1}$. 
 Extensions over the basic type graph $TG_{Base}$ of Fig.~\ref{fig:TG-Base} are colored
in light gray. 
 Controls \type{Job}, \type{Printer}, \type{Room}, \type{Spool}, \type{User} 
and \type{Computer} are modeled as node types, each of them being a subtype of 
the basic node type \type{BNode}. 
 Note that the well-formedness rule (\ref{eq:wfmd-arities}) which handles the 
arities of a signature is generic and needs not to be adjusted.
% Moreover, we get the following additional well-formedness rules properly 
% representing the arities defined by $\Sigma_1$:
\end{example}

\subsection{Mapping Bigraphs to Typed Graphs}
Now that we have specified how a bigraphical signature $\Sigma$ is mapped to a 
typed graph $TG_\Sigma$ in a compatible way, we can finally define the relation 
between a bigraph $B$ over $\Sigma$ and its corresponding typed graph 
$G$ over $TG_\Sigma$. 
 First, we define an \emph{element mapping} which bijectively maps the 
elements of $B$ to the nodes of $G$ in a suitable way. 
 Subsequently, a set of additional \emph{soundness criteria} will be 
specified over the basic element mapping.

\paragraph{Basic element mapping.}
Let us  first define a basic element mapping which bijectively maps the 
elements of a bigraph to the nodes of a typed graph in a suitable way.
 The domain of this mapping is the set of elements of a bigraph, which is the 
union of its nodes, edges, sites, roots, inner names, outer names and ports.
To establish this mapping formally,
we have to recall here that the identifiers of the sites and roots of a bigraph 
are drawn from potentially overlapping sets, namely the finite ordinals 
$\{0, \dots, k\}$ and $\{0, \dots, m\}$, respectively (see Section~\ref{sec:bigraphs}). 
Thus, in order to distinguish the roots and sites, we introduce two utility
functions, namely the two substitutions $\sigma_k : k \rightarrow k[\sigma_k]$ and $\sigma_m : m \rightarrow m[\sigma_m]$,
both injective mappings and we have $k[\sigma_k] \cap m[\sigma_m] = \emptyset$. 
% \begin{itemize}
%  \item $\sigma_k : k \rightarrow k[\sigma_k]$ is a substitution which injectively 
%        maps the finite ordinal $k$ onto a set $k[\sigma_k]$ disjoint from $k$.
%  \item $\sigma_m : m \rightarrow m[\sigma_m]$ is a substitution which injectively 
%        maps the finite ordinal $m$ onto a set $m[\sigma_m]$ disjoint from $m$
%        and disjoint from $k[\sigma_k]$.
% \end{itemize}
%
Analogously, identifiers for the inner names $X$ and the outer names $Y$ are drawn 
from the common infinite set $\mathcal{X}$ and thus potentially overlapping. 
Again, just to distinguish the inner names from the outer names, we introduce 
substitutions $\sigma_X : X \rightarrow X[\sigma_X]$ and $\sigma_Y : Y \rightarrow Y[\sigma_Y]$,
both injective mappings and we have $X[\sigma_X] \cap Y[\sigma_Y] = \emptyset$. 
% \begin{itemize}
%  \item $\sigma_X : X \rightarrow X[\sigma_X]$ is a substitution which injectively 
%        maps the finite ordinal $X$ onto a set $X[\sigma_X]$ disjoint from $X$.
%  \item $\sigma_Y : Y \rightarrow Y[\sigma_Y]$ is a substitution which injectively 
%        maps the finite ordinal $Y$ onto a set $Y[\sigma_Y]$ disjoint from $Y$
%        and disjoint from $X[\sigma_X]$.       
% \end{itemize}

We are now ready to define the mapping of the elements of a bigraph $B$ to the 
nodes of an instance graph $G$ as a bijective function
\begin{center}
$\phi : V_B \uplus E_B \uplus P_B \uplus k[\sigma_k] \uplus X[\sigma_X] \uplus m[\sigma_m] \uplus Y[\sigma_Y] \rightarrow G_N$ 
\end{center}
with
\begin{center}
$ \phi(x) = \left\{\begin{array}{cl} 
Id_{V_B}(x), & \mbox{if }x \in V_B\\ 
Id_{E_B}(x), & \mbox{if }x \in E_B\\ 
Id_{k[\sigma_k]}(x), & \mbox{if }x \in k[\sigma_k]\\
Id_{X[\sigma_X]}(x), & \mbox{if }x \in X[\sigma_X]\\
Id_{m[\sigma_m]}(x), & \mbox{if }x \in m[\sigma_m]\\
Id_{Y[\sigma_Y]}(x), & \mbox{if }x \in Y[\sigma_Y]\mbox{.}\\
\end{array}\right. $
\end{center}

\paragraph{Proper typing of nodes.}
The nodes in an instance graph $G$ must be properly typed 
to correctly represent a generic bigraph and the controls. 
Let $\Sigma$ be a signature and $TG_\Sigma$ its compatible type graph. 
Furthermore, let $B$ be a bigraph over $\Sigma$ and $G$ be an instance graph
over $TG_\Sigma$ with elements of $B$ bijectively mapped to the nodes of $G$
by the basic element mapping $\phi$. The nodes in $G$ are properly typed if
the following condition holds:

\begin{center}
$\begin{array}{lrcl}
   \forall x \in G_N : & & & \\
   & \bigl(\mbox{ }type_G(x) \in ran(\Phi)    & \implies & \phi^{-1}(x) \in V_B\mbox{ }\wedge\mbox{ }ctrl_B(\phi^{-1}(x)) = \Phi^{-1}(type_G(x))\mbox{ }\bigr)\mbox{ }\wedge \\
   & \bigl(\mbox{ }type_G(x) = BEdge          & \implies & \phi^{-1}(x) \in V_E\mbox{ }\bigr)\mbox{ }\wedge \\
   & \bigl(\mbox{ }type_G(x) = BSite          & \implies & \phi^{-1}(x) \in k[\sigma_k]\mbox{ }\bigr)\mbox{ }\wedge \\
   & \bigl(\mbox{ }type_G(x) = BInnerName     & \implies & \phi^{-1}(x) \in X[\sigma_X]\mbox{ }\bigr)\mbox{ }\wedge \\
   & \bigl(\mbox{ }type_G(x) = BRoot          & \implies & \phi^{-1}(x) \in m[\sigma_m]\mbox{ }\bigr)\mbox{ }\wedge \\
   & \bigl(\mbox{ }type_G(x) = BOuterName     & \implies & \phi^{-1}(x) \in Y[\sigma_Y]\mbox{ }\bigr) 
\end{array}$   
\end{center}

\paragraph{Additional soundness criteria.}
A bijective element mapping is necessary for 
inducing a unique transformation from bigraphs to typed graphs (and vice versa), 
however, it is obviously not sufficient.
 To that end, we introduce a set of additional soundness criteria defined over 
the basic element mapping satisfying typing constraints.
 In the following, let again $\Sigma$ be a signature and $TG_\Sigma$ its 
compatible type graph.
 Furthermore, let $B$ be a bigraph over $\Sigma$ and $G$ be an instance graph
over $TG_\Sigma$ with elements of $B$ bijectively mapped to the nodes of $G$
by the basic element mapping $\phi$, nodes of $G$ are properly typed. 
We have the following additional soundness criteria: %, where $dom(f)$ refers to the domain of a mapping $f$.

\begin{itemize}

\item[(1)] The nesting of places in $B$ must coincide with the containment structure in $G$:
\begin{center}
$\forall p_B \in V_B \uplus k :$ \\
    %p_B \in dom(prnt_B) \implies$ \\
	$\exists e_G \in G_E : \mbox{ }
	    src_G(e_G) = \phi(p_B) \mbox{ }\wedge\mbox{ }
	    tgt_G(e_G) =  \phi(prnt_B(p_B)) \mbox{ }\wedge\mbox{ } 
	    type_G(e_G) = bPrnt$		    
\end{center}

\begin{center}
$\forall e_G \in E_G : \mbox{ }
    type_G(e_G) = bPrnt \implies$ \\
	$\exists p_B \in V_B \uplus k : \mbox{ }
	    \phi(p_B) = src_G(e_G) \mbox{ }\wedge\mbox{ }
	    \phi(prnt_B(p_B)) = tgt_G(e_G)$
\end{center}

\item[(2)] The linking structure in $B$ must coincide with the linking structure in $G$:
\begin{center}
$\forall p_B \in X \uplus P_B :$ \\
   % p_B \in dom(link_B) \implies$ \\
	$\exists e_G \in G_E : \mbox{ }
	    src_G(e_G) = \phi(p_B) \mbox{ }\wedge\mbox{ }
	    tgt_G(e_G) =  \phi(link_B(p_B)) \mbox{ }\wedge\mbox{ } 
	    type_G(e_G) = bLink$	
\end{center}

\begin{center}
$\forall e_G \in E_G : \mbox{ }
    type_G(e_G) = bLink \implies$ \\
	$\exists p_B \in X \uplus P_B : \mbox{ }
	    \phi(p_B) = src_G(e_G) \mbox{ }\wedge\mbox{ }
	    \phi(link_B(p_B)) = tgt_G(e_G)$
\end{center}

\item[(3)] The indexing of the roots in $B$ must be consistent to
the indices of their corresponding representations in $G$:

\begin{center}
$\forall i \in \{0, \dots, m-1\}, \mbox{ } n_G \in G_N \mbox{ } : \mbox{ }
    (i, n_G) \in \phi \mbox{ }\Longleftrightarrow\mbox{ } n_G.index = i$ 	    
\end{center}

\item[(4)] The indexing of the sites in $B$ must be consistent to
the indices of their corresponding representations in $G$:

\begin{center}
$\forall i \in \{0, \dots, k-1\}, \mbox{ } n_G \in G_N \mbox{ } : \mbox{ }
    (i, n_G) \in \phi \mbox{ }\Longleftrightarrow\mbox{ } n_G.index = i$ 	    
\end{center}

\item[(5)] The indexing of the ports in $B$ must be consistent to
the indices of their corresponding representations in $G$:

\begin{center}
$\forall (v,i) \in P_B, \mbox{ } n_G \in G_N \mbox{ } : \mbox{ }
    ((v,i), n_G) \in \phi \mbox{ }\Longleftrightarrow\mbox{ } n_G.index = i$ 	    
\end{center}

\end{itemize}

\begin{figure}[ht]
 \centering
 \includegraphics[width=0.95\textwidth]{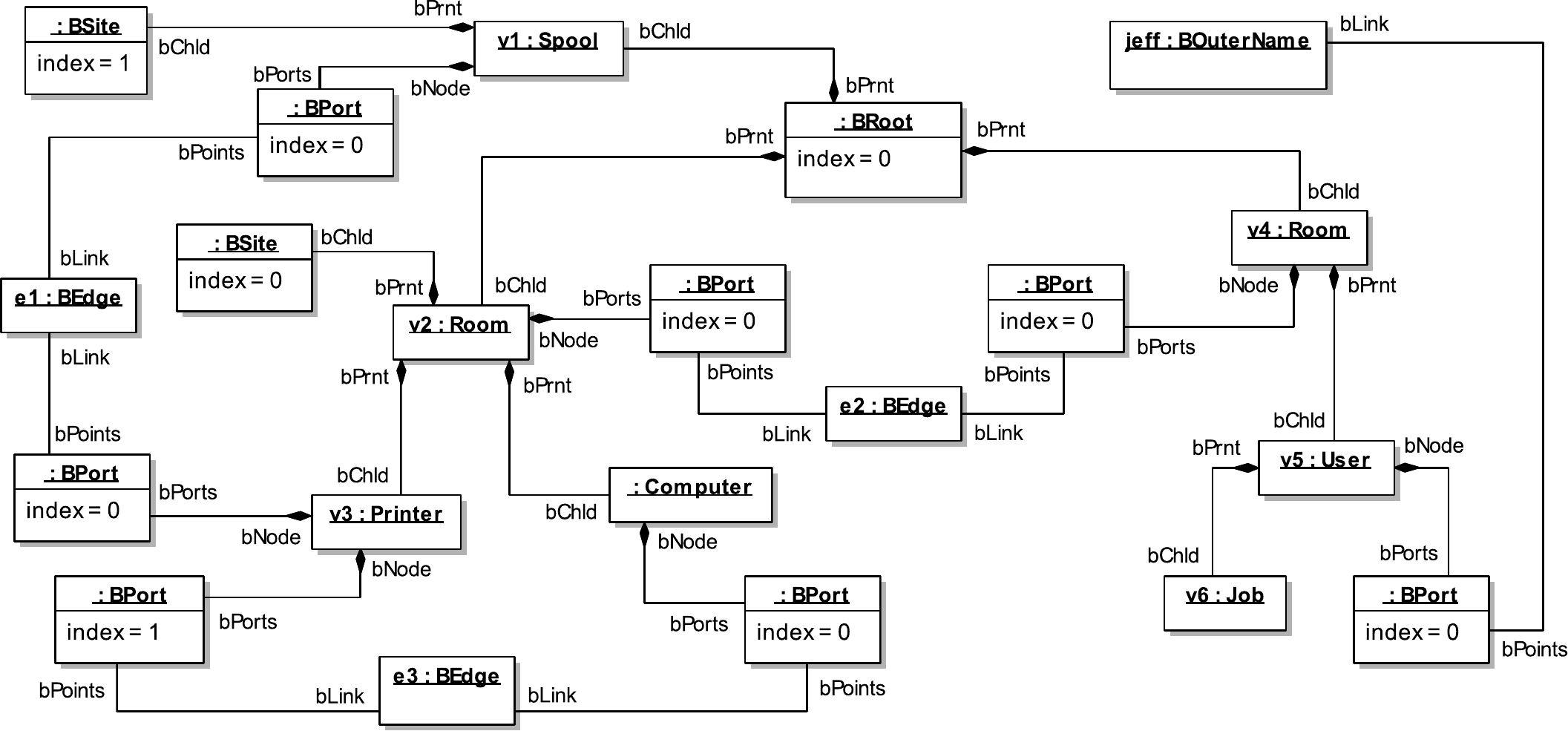}
 \caption{Sample bigraph $B_1$ over $\Sigma_1$ as typed graph $G_1$ over $TG_{\Sigma_1}$}
 \label{fig:G-Printer}
\end{figure}

\begin{example}[Sample bigraph as typed graph]\label{ex:printer-graph}
The sample bigraph $B_1$ over 
$\Sigma_1$ of Fig.~\ref{fig:printer} is mapped to the graph $G_1$ typed over the 
printing system type graph $TG_{\Sigma_1}$ shown in Fig.~\ref{fig:G-Printer}.
 Note that some of the nodes of Fig.~\ref{fig:G-Printer} have 
identifiers, indicated by the notation \elmt{id:NodeType}. They are aligned
to the identifiers of bigraphical nodes and edges used in Fig.~\ref{fig:printer}
and thus symbolize parts of the bijective mapping $\phi$ from bigraph elements 
in $B_1$ to the nodes in $G_1$.
 For the remaining nodes, identifiers are omitted in Fig.~\ref{fig:G-Printer}.
%The mapping to bigraph elements of $B_1$ is clear by the respective context.
All nodes are properly typed and additional soundness criteria regarding the 
nesting of places, the linking structure as well as the indexing of roots,
sites and ports are fulfilled.
\end{example}

%% file: variations.tex
\section{Application-Specific Variation Points}
\label{sec:variations}

In general, the definition of an appropriate abstract syntax of a modeling 
language depends on several design decisions, many of which are 
application-specific. 
 The standard UML meta-model, for example, 
mainly targets the construction of UML editors but has several drawbacks for the 
construction of model differencing tools, see e.g.\ \cite{kehrer2011lifting}.
 In this section, we address application-specific variation points of an abstract 
syntax for bigraphs. 
 In order to deal with this variability, we apply basic principles known from the 
field of software product line engineering (SPLE) to optimally support a wide 
range of applications. 
 In Section~\ref{sec:problem-space}, we first identify variation points of an abstract syntax for 
bigraphs. The results of our variability analysis are formally documented in 
terms of a feature model yielding the configuration space of a bigraphical
abstract syntax.
 Subsequently, Section~\ref{sec:solution-space} briefly discusses how to reconfigure canonical 
type and instance graphs using standard variability mechanisms adopted from
model-based software product line engineering. 
 We do not claim the presented variability model to be complete; however, the proposed
framework can be easily extended to support additional variation points and variants.

% derived by our canoncical mapping from bigraphs to typed 
% graphs presented in Section~\ref{}.
% 
%  implement and bind the 
% identified variation points using standard methods for engineering model-based 
% software product lines. 

\subsection{Variability Model}
\label{sec:problem-space}
\begin{figure}[ht]
 \centering
 \includegraphics[width=0.9\textwidth]{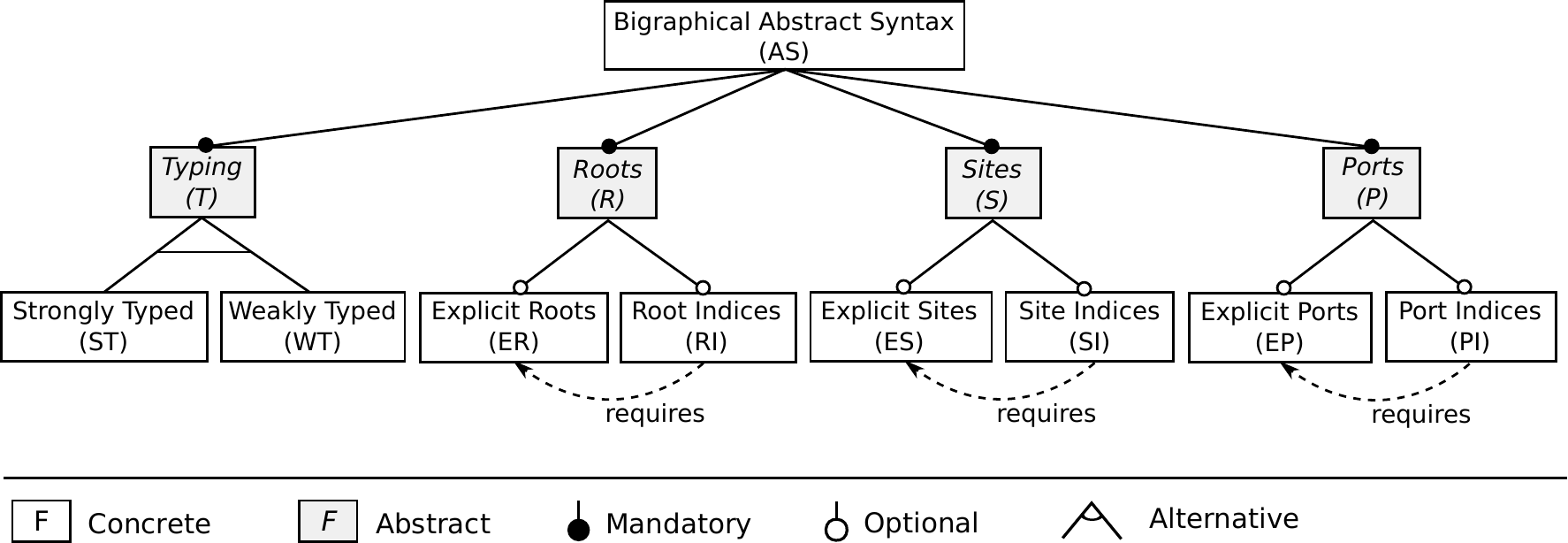}
 \caption{Variability model defining common and variable parts among possible 
variants of a bigraphical abstract syntax}
 \label{fig:FeatureModel}
\end{figure}

Feature models are a widely used method to model variability in software product
line engineering. FODA-like feature models~\cite{FODA} have an intuitive 
tree-like graphical syntax and a precise formal semantics which can be denoted 
as propositional formulae over Boolean feature variables~\cite{BatorySPLC2005}.
 The features of Fig.~\ref{fig:FeatureModel} define common and variable parts among possible 
variants of a bigraphical abstract syntax. 
 The root feature, called \feature{Bigraphical Abstract Syntax (AS)}, has four 
sub-features constituting the main variation points. 
Please note that all of these four features are abstract features (the feature 
names are printed in italics in the feature diagram), i.e. they are merely used 
to structure the feature model along the main variation points.

\MDB{mm}{\em}

\item[Typing:]
The variation point \feature{Typing (T)} defines two alternatives. In the 
\feature{Strongly Typed (ST)} variant, the types of graph nodes representing 
bigraphical nodes are defined by the typing morphism. This variant is selected 
for our canonical definition of an abstract syntax presented in
Section~\ref{sec:mapping}.
In a \feature{Weakly Typed (WT)} variant, controls of bigraphical nodes are 
simply attached as attributes to the corresponding nodes in the typed graph,
while the nodes themselves are of the generic node type \type{BNode}.
%This variant allows bigraphical nodes to be dynamically retyped by an application.
Weak typing is sensible if applications shall dynamically retype nodes 
representing bigraphical nodes.

\item[Roots:]
Concerning the variation point \feature{Roots (R)}, our canonical definition of 
an abstract syntax for bigraphs comprises the two optional sub-features 
\feature{ER} and \feature{RI}.
 Firstly, the feature \feature{Explicit Roots (ER)} denotes that bigraphical roots are 
explicitly represented as dedicated nodes in the typed graph. 
 Secondly, feature \feature{Root Indices (RI)} means that 
bigraphical roots are indexed in the graph-based representation, which requires 
that bigraphical roots are explicitly represented.
 The representation of root indices is necessary if a graph-based representation
needs to be re-transformed into a bigraph without the loss of any information.
 However, if this is not required in an application context and if we are
striving for a compact representation, optional features \feature{RI} and \feature{ER} 
can be omitted. 
% 
% root nodes in our graph-based representation can be derived; 
% a node which has no parent node, i.e.\ no outgoing edge of type \type{bPrnt},
% represents a bigraphical root.
%  Our canonical definition of an abstract syntax for bigraphs comprises both
% optional features \feature{ER} and \feature{RI}.

\item[Sites and ports:]
Similar variants like the one discussed for the variation point \feature{Roots (R)} 
exist for the graph-based representation of sites and ports, 
cf.\ variation points \feature{Sites (S)} and \feature{Ports (P)} in 
Fig.~\ref{fig:FeatureModel}.
Likewise, all optional sub-features, namely 
\feature{Explicit Sites (ES)} and \feature{Site Indices (SI)}
as well as
\feature{Explicit Ports (ES)} and \feature{Port Indices (PI)}
are present in the canonical abstract syntax.

\MDE

\subsection{Reconfiguring the Canonical Abstract Syntax}
\label{sec:solution-space}
To reconfigure the canonical abstract syntax for bigraphs according to a
selected feature configuration, suitable variability mechanisms
have to be chosen on the type and on the instance level, i.e.\ both type and 
instance graphs have to be configured according to a selected feature 
combination.

% As usual in software product line engineering, the features defined by our 
% feature model of Figure~\ref{FeatureModel} represent the characteristics of
% different variants of an abstract syntax for bigraphs on an abstract level.
% In order to adapt the representation according to a set of selected 
% features, all features have to be mapped to concrete model fragments which are 
% bound to a specific variant upon request. Here, suitable variability mechanisms
% have to be chosen on the type and on the instance level, i.e.\ both type and 
% instance graphs have to be configured according to a selected feature 
% combination.

\begin{figure}[ht]
 \centering
 \includegraphics[width=1.0\textwidth]{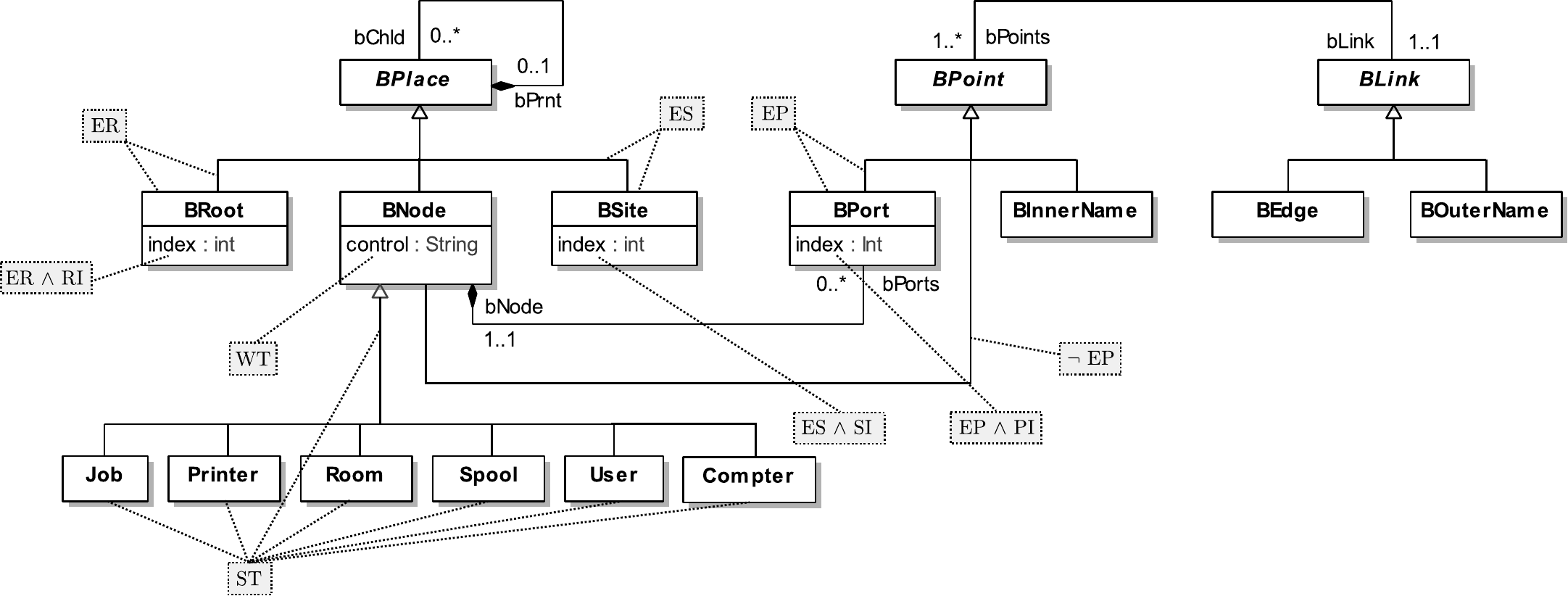}
 \caption{150\% meta-model for our running printer example}
 \label{fig:150}
\end{figure}

\paragraph{Type-level variability.}
We adopt the idea of model templates~\cite{czarnecki2005mapping}, also known as 150\% models, to establish a suitable 
variability mechanism on the type level; instead of having a fixed type graph 
derived from a bigraphical signature, we use a 150\% type graph which is a 
superimposition of all possible variants and includes variability.
%in the specification of an abstract syntax for a bigraphical modeling language. 
A 150\% type graph for our running printer example
%, in the following referred to as $TG_{\Sigma_1,150}$, 
is shown in Fig.~\ref{fig:150}.
It differs from the canonical type graph of Fig.~\ref{fig:G-Printer} in that \type{BNode}
has an additional attribute \type{control:String} and is defined as a subtype
of \type{BPoint}.
 Moreover, some of the elements of the 150\% type graph are annotated by 
propositional formulae over features, thus implicitly defining a mapping between 
features and the 150\% type graph.
 Given a desired configuration in terms of a valid feature combination, variability 
is resolved as follows:
%\begin{itemize}
  Elements which are not mapped to a feature are included in any possible configuration;
  variable elements are included in a configuration if the annotated 
       propositional formula over features evaluates to \emph{true}.
%\end{itemize}
For example, the subtype relationship between \type{BNode} and 
\type{BPoint} will only be included in the final ``100\% type graph'' if ports 
shall be not represented explicitly ($\neg$\feature{EP}), i.e.\ nodes of type
\type{BNode} can be linked to nodes of types \type{BEdge} and \type{BOuterName}
directly in that case.
%  Furthermore, the additional attribute \type{control:String}
% will be included if a weakly typing (\feature{WT}) is chosen,
% otherwise controls are modeled as node types explcitly (\feature{WT}).

Extending the canonical type graph obtained for a given bigraphical signature 
to a 150\% type graph is straightforward. 
 Most of the extensions, including the respective annotations, 
refer to the basic type graph and can be handled in a generic way. 
 In addition, annotation \feature{ST} refers to each node type and the 
respective subtyping relationship induced by controls of the signature.

\paragraph{Instance-level variability.}
At the instance level, we use a delta-oriented approach~\cite{delta1,delta2} 
as a suitable 
variability mechanism. The idea is that the canonical graph representation 
derived from a given bigraph, which can be referred to as core variant in the 
delta-oriented terminology, is transformed into an alternative variant by the 
application of a set of deltas. A delta is basically a patch removing, replacing 
or adding instance graph elements related to a feature. Moreover, a delta is
equipped with a delta application condition. 
Similarly to the annotation in the 150\% type graph, a delta application
condition is a propositional expression over features of the feature model.
A delta has to be applied to the core variant if the propositional formula
evaluates to \emph{true}. 
An overview of the required deltas, their application conditions (AC) and a 
brief description of each delta is presented in Table~\ref{tab:deltas}. 
% We assume here that the transformation specified by a delta can be performed
% based on the 150\% type graph.

\begin{table}[h!]
\centering

{\small
\begin{tabular}{ p{2cm} | p{1cm} | p{11.5cm} }
                  & {\bf AC}           & {\bf Description} \\
  \hline\hline 
  $\Delta(ST,WT)$ & \feature{WT}       & To switch from the strongly typed variant to the the weakly typed one, each node representing a bigraphical node has to be retyped to the generic node type \type{BNode} and the value of the attribute \elmt{control} has to be set accordingly.\\
  \hline
  $\Delta(RI,ER)$ & $\neg$\feature{RI} & To remove the indexing of root nodes, the attribute \elmt{index} has to be unset for each node of type \type{BRoot}. \\
  $\Delta(ER,AS)$ & $\neg$\feature{ER} & To abstain from an explicit representation of root nodes, each node of type \type{BRoot} together with its incident edges has to be deleted.\\
  \hline
  $\Delta(SI,ES)$ & $\neg$\feature{SI} & Analogous to $\Delta(RI,ER)$. \\
  $\Delta(ES,AS)$ & $\neg$\feature{ES} & Analogous to $\Delta(ER,AS)$. \\
  \hline 
  $\Delta(PI,EP)$ & $\neg$\feature{PI} & Analogous to $\Delta(RI,ER)$. \\
  $\Delta(EP,AS)$ & $\neg$\feature{EP} & Analogous to $\Delta(ER,AS)$. \\
\end{tabular}
}

\caption{Set of deltas specifying basic reconfigurations on the instance level}
\label{tab:deltas}
\end{table}

%% file: tooling.tex
\section{Tool Integration and Example Application}
\label{sec:tooling}

In this section, we briefly outline how to integrate the approach into an
existing MDE tool environment. We give an example application of such an 
integration showcasing the benefits of the proposed approach.

\paragraph{Integration into an MDE environment.}
Type and instance graphs can be implemented as meta-models and models using 
various technical MDE frameworks.
 In principle, any framework can be used as long as it supports the EMOF 
standard defined by the OMG. 
%  An overview of how to integrate the approach presented in this paper into an 
% MDE tool environment is illustrated in Figure~\ref{}.\todo{do we need that?} 
 Since the available bigraphical research prototypes use tool-specific 
representations, input and output transformations have to be developed for 
interfacing with them. The mapping of bigraphs to typed graphs presented in Section \ref{sec:mapping} serves as
a formal specification for this.
 The variability framework presented in Section \ref{sec:variations} %, notably the required reconfigurations of canonical meta-models and models, 
 can be
implemented using available model transformation techniques.
 As a proof of concept, we implemented the approach as an extension of the 
Eclipse IDE integrating the bigraph modeling tool Big Red~\cite{faithfull2013big} 
with the Eclipse Modeling Framework \cite{EMFBook}, the de-facto reference
implementation of OMG's EMOF standard.

\paragraph{Example application.}
% Having input and output transformations as sketched in Figure~\ref{} at hand, 
% tool chaining bigraphical modeling and analyis tools as well as integrating 
% them with standard MDE tools is straightforward. 
As an example application, we extended the bigraph modeling environment 
Big Red by an additional constraint checking facility which is based on the 
Object Constraint Language (OCL)~\cite{OCL}.
% This integration is motivated by the fact that, besides the generic bigraphical 
% anatomy and the arities of controls, the definition of additional well-formedness 
% rules restricting the set of meaningful models for a bigraphical modeling language
% is not supported.
This extension is motivated by the fact that the designer of a bigraphical domain-specific modeling language
might wish to add context-sensitive constraints to the language definition
restricting the set of meaningful models for the domain of interest and supporting automated constraint validation.
In our running example of a context-aware printer system, we can think of the following invariants:
% In Example~\ref{}, we have introduced domain-specific bigraphical language for
% modeling static aspects. In Big Red, the visual syntax of this DSL can be
% further customized by assigning user-defined graphical shapes to controls
% of the signature. 

\begin{itemize}
 \item[(iv1)] A \type{Spool} may only contain \type{Jobs} and \type{Sites} as nested places. 
 \item[(iv2)] Due to capacity restrictions, a \type{Spool} may contain at most
              a pre-defined number $n$ of Jobs. Moreover, if the maximum capacity 
              of $n$ \type{Jobs} is reached, the \type{Spool} may not contain a \type{Site} representing 
              further unspecified \type{Jobs}.
 \item[(iv3)] \type{Rooms} may be linked to each other only, i.e.\ they must not be 
              linked to other nodes or outer names.
\end{itemize}

In MDE, invariants like these are typically specified using OCL expressions defined over the meta-model and used to reason on instance structures.
Having an EMOF-compliant abstract syntax for bigraphs at hand, we can 
formally capture our sample invariants in OCL as shown below. We assume OCL
expressions to be defined over the canonical meta-model of the printer example,
and the capacity $n$ of the \type{Spool} (iv2) to be set to 100 \type{Jobs}. 
Existing work on bigraphs does not provide a dedicated concept to express 
invariants like these. 
As a workaround, some invariants could be specified as bigraph
patterns which are forbidden and use bigraph matching to verify their absence, however, very
inconveniently; for invariant iv1, for instance, a bigraphical anti-pattern would
consist of a \type{Spool} containing some node having a control different from \type{Job}. 
An anti-pattern is required for all remaining controls in the signature.

\begin{figure*}[t]
\vspace{0.25cm}
{\small
\noindent
{\bf context} Spool\\
\mbox{ }\hspace{0.25cm}{\bf inv} iv1:\\ 
\mbox{ }\hspace{0.5cm}{\bf self}.bChld--$>$forAll(c $\mbox{}\mid$ c.oclIsTypeOf(BSite) {\bf or} c.oclIsTypeOf(Job))\\
\mbox{ }\hspace{0.25cm}{\bf inv} iv2:\\
\mbox{ }\hspace{0.5cm}{\bf let} n : integer = 100\\
\mbox{ }\hspace{0.5cm}{\bf self}.bChld--$>$size() $<$= n {\bf and}\\
\mbox{ }\hspace{0.75cm}{\bf self}.bChld--$>$size() = n {\bf implies} {\bf not}(self.bChld--$>$exists( c $\mbox{}\mid$ c.oclIsTypeOf(BSite)))\\
{\bf context} Room\\
\mbox{ }\hspace{0.25cm}{\bf inv} iv3:\\
\mbox{ }\hspace{0.5cm}{\bf let} port : BPort = {\bf self}.bPorts--$>$first()\\
\mbox{ }\hspace{0.5cm}port.bLink.oclIsTypeOf(BEdge) {\bf and} port.bLink.bPoints--$>$forAll(\\
\mbox{ }\hspace{0.75cm}p $\mbox{}\mid$ p.oclIsTypeOf(BPort) {\bf and} p.oclAsType(BPort).bNode.oclIsTypeOf(Room)))\\
}
% 
%  \caption{Sample bigraph $B_1$ over $\Sigma_1$ as typed graph $G_1$ over $TG_{\Sigma_1}$}
%  \label{fig:G-Printer}
\vspace{-0.5cm}
\end{figure*}

% \lstset{language=OCL}
% \begin{lstlisting}[float,floatplacement=t,label=listing:signature, caption=TODO, mathescape=true]
% context Spool
%   inv iv1: 
%     self.bChld->forAll(c | c.oclIsTypeOf(BSite) or c.oclIsTypeOf(Job))
%   inv iv2:
%     let n : integer = 100
%     self.bChld->size() <= n and
%       self.bChld->size() = n implies not(self.bChld->exists(c | c.oclIsTypeOf(BSite)))
% context Room
%   inv iv3:
%     let port : BPort = self.bPorts->first()
%     self.port.bLinks->forAll(
%       l | l.oclIsTypeOf(BEdge) and l.bPoints->forAll(
% 	p | p.oclIsTypeOf(BPort) and p.oclAsType(BPort).bNode.oclIsTypeOf(Room))
% \end{lstlisting}

% 
% 
% \todo[inline]{The value is in showing that we can easily express constraints on the structures without resorting to bg matching. 
%    Should we mention here that constraints can be also expressed as bigraphical anti-patterns, however, very often quite inconveniently?
%    Consider for example iv1; here, an anti-pattern specifies a Job containing some node having a control different from Job. This has to be done for all controls in the signature.
%    Similarly, consider example iv2; here, an anti-pattern specifies a Spool containing more than $n$ jobs, which can be many...}
% \todo[inline]{Maybe we should mention that at least the theory provides the 
% sorting as a means to define some kind of constraints?}

%% file: conclusion.tex
%!TEX root = ./paper.tex

\section{Conclusion and Future Work}
\label{sec:conclusion}

% Bigraphs are an emerging formalism and domain-specific language for designing systems in ubiquitous computing. 
%  They enjoy an intuitive and at the same time unambiguous visual syntax but yet
% lack a precise definition of an abstract syntax.

% In this paper, we reported on ongoing work to create software tools that will 
% bring bigraphs to life as a language for programming and simulation.
% \todo[inline]{Mirror somehow the introduction and discuss open issues and topics for future work}

In this paper, we presented an abstract syntax for bigraphs and reported on ongoing work to
realize a complete framework for support of bigraph-enabled reasoning through state of the art model-driven engineering techniques.
Our effort is motivated by the lack of a precise definition of an abstract syntax for 
bigraphical modeling languages, which hinders the interoperability between existing bigraphical tools, 
and the need to enable development support, facilitate reasoning and evaluate the bigraphical theory in real-world applications through MDE technologies and modeling environments.
We proposed an abstract syntax for bigraphs which is compliant with the EMOF standard defined by the OMG. We used typed graphs as a formal underpinning of EMOF-based models and presented a canonical mapping from bigraphs to typed graphs. 
Since the abstract syntax of a modeling language depends on application-specific design decisions, we discussed variation points 
in the graph-based representation of bigraphs and showed how variability can be handled based on 
software product line techniques.

%In this paper, we have focussed on static aspects of the bigraphical theory.
As for future work, to realize a complete framework, we will need to better understand the relation between bigraphical reactive systems and 
graph transformation systems~\cite{Ehrig2006}.
%, notably the algebraic approach to graph transformation~\cite{Ehrig2006}.
%Our hypothesis is that a bigraphical reaction rule can be expressed as a graph transformation rule if certain conditions between redex and reactum hold, such as that the mapping of reactum sites to redex sites is bijective.
Moreover, the mapping previously presented can be extended to other forms of bigraphs, e.g.\ bigraphs with sharing~\cite{sevegnani2015bigraphs}, with directed edges~\cite{bacci2009dbtk} or stochastic semantics~\cite{hojsgaard2012bigraphical}.
From a technical point of view, we plan to provide a reference implementation based on Eclipse to the community, facilitating interchange of bigraphical models and enabling the integration of MDE-based techniques.

% \todo[inline]{Future Work TBD, here are some possible points:}
% \begin{itemize}
%  \item Extend the mapping to other forms of bigraphs, e.g.\ bigraphs with sharing as proposed by Sevegnany.
%  \item Simulating bigraphical reactions as graph transformations. 
%        More generally, study the relation between BRS and graph transformation systems.
%        Our hypothesis is that a bigraphical rule can be expressed as graph transformation rule if mapping of reactum sites to redex sites is bijective.
%  \item From a technical point of view: Provide the reference implementation based on Eclipse to the community, e.g.\ as project on GitHub
% \end{itemize}

%% file: main.bbl
\begin{thebibliography}{10}
\providecommand{\bibitemdeclare}[2]{}
\providecommand{\surnamestart}{}
\providecommand{\surnameend}{}
\providecommand{\urlprefix}{Available at }
\providecommand{\url}[1]{\texttt{#1}}
\providecommand{\href}[2]{\texttt{#2}}
\providecommand{\urlalt}[2]{\href{#1}{#2}}
\providecommand{\doi}[1]{doi:\urlalt{http://dx.doi.org/#1}{#1}}
\providecommand{\bibinfo}[2]{#2}

\bibitemdeclare{incollection}{bacci2009dbtk}
\bibitem{bacci2009dbtk}
\bibinfo{author}{Giorgio \surnamestart Bacci\surnameend},
  \bibinfo{author}{Davide \surnamestart Grohmann\surnameend} \&
  \bibinfo{author}{Marino \surnamestart Miculan\surnameend}
  (\bibinfo{year}{2009}): \emph{\bibinfo{title}{{DBtk: A Toolkit for Directed
  Bigraphs}}}.
\newblock In: {\sl \bibinfo{booktitle}{Proc.\ Intl.\ Conf.\ on Algebra and
  Coalgebra in Computer Science}}, \bibinfo{publisher}{Springer}, pp.
  \bibinfo{pages}{413--422}, \doi{10.1007/978-3-642-03741-2\_28}.

\bibitemdeclare{inproceedings}{BatorySPLC2005}
\bibitem{BatorySPLC2005}
\bibinfo{author}{Don \surnamestart Batory\surnameend} (\bibinfo{year}{2005}):
  \emph{\bibinfo{title}{Feature Models, Grammars, and Propositional Formulas}}.
\newblock In: {\sl \bibinfo{booktitle}{Proc.\ Intl.\ Conf.\ on Software Product
  Lines}}, \bibinfo{publisher}{Springer}, pp. \bibinfo{pages}{7--20},
  \doi{10.1007/11554844\_3}.

\bibitemdeclare{article}{biermann2012formal}
\bibitem{biermann2012formal}
\bibinfo{author}{Enrico \surnamestart Biermann\surnameend},
  \bibinfo{author}{Claudia \surnamestart Ermel\surnameend} \&
  \bibinfo{author}{Gabriele \surnamestart Taentzer\surnameend}
  (\bibinfo{year}{2012}): \emph{\bibinfo{title}{Formal foundation of consistent
  EMF model transformations by algebraic graph transformation}}.
\newblock {\sl \bibinfo{journal}{Software \& Systems Modeling}}
  \bibinfo{volume}{11}(\bibinfo{number}{2}), pp. \bibinfo{pages}{227--250},
  \doi{10.1007/s10270-011-0199-7}.

\bibitemdeclare{inproceedings}{bundgaard2006typed}
\bibitem{bundgaard2006typed}
\bibinfo{author}{Mikkel \surnamestart Bundgaard\surnameend} \&
  \bibinfo{author}{Vladimiro \surnamestart Sassone\surnameend}
  (\bibinfo{year}{2006}): \emph{\bibinfo{title}{Typed polyadic pi-calculus in
  bigraphs}}.
\newblock In: {\sl \bibinfo{booktitle}{Proc.\ Intl.\ Conf.\ on Principles and
  Practice of Declarative Programming}}, \bibinfo{organization}{ACM}, pp.
  \bibinfo{pages}{1--12}, \doi{10.1145/1140335.1140336}.

\bibitemdeclare{inproceedings}{Cardelli98mobileambients}
\bibitem{Cardelli98mobileambients}
\bibinfo{author}{Luca \surnamestart Cardelli\surnameend} \&
  \bibinfo{author}{Andrew~D.\ \surnamestart Gordon\surnameend}
  (\bibinfo{year}{1998}): \emph{\bibinfo{title}{{Mobile Ambients}}}.
\newblock In: {\sl \bibinfo{booktitle}{Proc.\ Intl.\ Conf.\ on Foundations of
  Software Science and Computation Structure}}, pp. \bibinfo{pages}{140--155},
  \doi{10.1007/BFb0053547}.

\bibitemdeclare{article}{delta1}
\bibitem{delta1}
\bibinfo{author}{Dave \surnamestart Clarke\surnameend},
  \bibinfo{author}{Michiel \surnamestart Helvensteijn\surnameend} \&
  \bibinfo{author}{Ina \surnamestart Schaefer\surnameend}
  (\bibinfo{year}{2015}): \emph{\bibinfo{title}{Abstract delta modelling}}.
\newblock {\sl \bibinfo{journal}{Mathematical Structures in Computer Science}}
  \bibinfo{volume}{25}(\bibinfo{number}{3}), pp. \bibinfo{pages}{482--527},
  \doi{10.1017/S0960129512000941}.

\bibitemdeclare{inproceedings}{czarnecki2005mapping}
\bibitem{czarnecki2005mapping}
\bibinfo{author}{Krzysztof \surnamestart Czarnecki\surnameend} \&
  \bibinfo{author}{Micha{\l} \surnamestart Antkiewicz\surnameend}
  (\bibinfo{year}{2005}): \emph{\bibinfo{title}{Mapping features to models: A
  template approach based on superimposed variants}}.
\newblock In: {\sl \bibinfo{booktitle}{Proc.\ Intl.\ Conf.\ on Generative
  Programming and Component Engineering}}, \bibinfo{organization}{Springer},
  pp. \bibinfo{pages}{422--437}, \doi{10.1007/11561347\_28}.

\bibitemdeclare{book}{Ehrig2006}
\bibitem{Ehrig2006}
\bibinfo{author}{Hartmut \surnamestart Ehrig\surnameend},
  \bibinfo{author}{Karsten \surnamestart Ehrig\surnameend},
  \bibinfo{author}{Ulrike \surnamestart Prange\surnameend} \&
  \bibinfo{author}{Gabriele \surnamestart Taentzer\surnameend}
  (\bibinfo{year}{2006}): \emph{\bibinfo{title}{Fundamentals of Algebraic Graph
  Transformation}}.
\newblock \bibinfo{publisher}{Springer}, \doi{10.1007/3-540-31188-2}.

\bibitemdeclare{article}{ehrig2009generating}
\bibitem{ehrig2009generating}
\bibinfo{author}{Karsten \surnamestart Ehrig\surnameend},
  \bibinfo{author}{Jochen~Malte \surnamestart K{\"u}ster\surnameend} \&
  \bibinfo{author}{Gabriele \surnamestart Taentzer\surnameend}
  (\bibinfo{year}{2009}): \emph{\bibinfo{title}{Generating instance models from
  meta models}}.
\newblock {\sl \bibinfo{journal}{Software \& Systems Modeling}}
  \bibinfo{volume}{8}(\bibinfo{number}{4}), pp. \bibinfo{pages}{479--500},
  \doi{10.1007/s10270-008-0095-y}.

\bibitemdeclare{article}{faithfull2013big}
\bibitem{faithfull2013big}
\bibinfo{author}{Alexander~John \surnamestart Faithfull\surnameend},
  \bibinfo{author}{Gian \surnamestart Perrone\surnameend} \&
  \bibinfo{author}{Thomas~T. \surnamestart Hildebrandt\surnameend}
  (\bibinfo{year}{2013}): \emph{\bibinfo{title}{Big Red: {A} Development
  Environment for Bigraphs}}.
\newblock {\sl \bibinfo{journal}{{ECEASST}}} \bibinfo{volume}{61},
  \doi{10.14279/tuj.eceasst.61.835}.

\bibitemdeclare{misc}{MOF}
\bibitem{MOF}
\bibinfo{author}{Object~Management \surnamestart Group\surnameend}
  (\bibinfo{year}{2013}): \emph{\bibinfo{title}{Meta Object Facility (MOF) Core
  Specification, version 2.4.1}}.
\newblock \bibinfo{howpublished}{OMG document number: formal/2013-06-01}.

\bibitemdeclare{misc}{OCL}
\bibitem{OCL}
\bibinfo{author}{Object~Management \surnamestart Group\surnameend}
  (\bibinfo{year}{2014}): \emph{\bibinfo{title}{Object Constraint Language
  (OCL), version 2.4}}.
\newblock \bibinfo{howpublished}{OMG document number: formal/2014-02-03}.

\bibitemdeclare{inproceedings}{Heckel2002Attributed}
\bibitem{Heckel2002Attributed}
\bibinfo{author}{Reiko \surnamestart Heckel\surnameend},
  \bibinfo{author}{Jochen~Malte \surnamestart K{\"{u}}ster\surnameend} \&
  \bibinfo{author}{Gabriele \surnamestart Taentzer\surnameend}
  (\bibinfo{year}{2002}): \emph{\bibinfo{title}{Confluence of Typed Attributed
  Graph Transformation Systems}}.
\newblock In: {\sl \bibinfo{booktitle}{Proc.\ Intl.\ Conf.\ on Graph
  Transformation}}, {\sl \bibinfo{series}{Lecture Notes in Computer Science}}
  \bibinfo{volume}{2505}, \bibinfo{publisher}{Springer}, pp.
  \bibinfo{pages}{161--176}, \doi{10.1007/3-540-45832-8\_14}.

\bibitemdeclare{phdthesis}{hojsgaard2012bigraphical}
\bibitem{hojsgaard2012bigraphical}
\bibinfo{author}{Espen \surnamestart H{\o}jsgaard\surnameend}
  (\bibinfo{year}{2012}): \emph{\bibinfo{title}{Bigraphical languages and their
  simulation}}.
\newblock Ph.D. thesis, \bibinfo{school}{IT University of Copenhagen}.

\bibitemdeclare{techreport}{FODA}
\bibitem{FODA}
\bibinfo{author}{Kyo \surnamestart Kang\surnameend}, \bibinfo{author}{Sholom
  \surnamestart Cohen\surnameend}, \bibinfo{author}{James \surnamestart
  Hess\surnameend}, \bibinfo{author}{William \surnamestart Novak\surnameend} \&
  \bibinfo{author}{A.~\surnamestart Peterson\surnameend}
  (\bibinfo{year}{1990}): \emph{\bibinfo{title}{Feature-Oriented Domain
  Analysis (FODA) Feasibility Study}}.
\newblock \bibinfo{type}{Technical Report} \bibinfo{number}{CMU/SEI-90-TR-021},
  \bibinfo{institution}{Software Engineering Institute, Carnegie Mellon
  University}, \bibinfo{address}{Pittsburgh, PA}.

\bibitemdeclare{inproceedings}{kehrer2011lifting}
\bibitem{kehrer2011lifting}
\bibinfo{author}{Timo \surnamestart Kehrer\surnameend}, \bibinfo{author}{Udo
  \surnamestart Kelter\surnameend} \& \bibinfo{author}{Gabriele \surnamestart
  Taentzer\surnameend} (\bibinfo{year}{2011}): \emph{\bibinfo{title}{A
  rule-based approach to the semantic lifting of model differences in the
  context of model versioning}}.
\newblock In: {\sl \bibinfo{booktitle}{Proc.\ Intl.\ Conf.\ on Automated
  Software Engineering}}, pp. \bibinfo{pages}{163--172},
  \doi{10.1109/ASE.2011.6100050}.

\bibitemdeclare{article}{krivine2008stochastic}
\bibitem{krivine2008stochastic}
\bibinfo{author}{Jean \surnamestart Krivine\surnameend}, \bibinfo{author}{Robin
  \surnamestart Milner\surnameend} \& \bibinfo{author}{Angelo \surnamestart
  Troina\surnameend} (\bibinfo{year}{2008}): \emph{\bibinfo{title}{Stochastic
  bigraphs}}.
\newblock {\sl \bibinfo{journal}{Electronic Notes in Theoretical Computer
  Science}} \bibinfo{volume}{218}, pp. \bibinfo{pages}{73--96},
  \doi{10.1016/j.entcs.2008.10.006}.

\bibitemdeclare{article}{leifer2006transition}
\bibitem{leifer2006transition}
\bibinfo{author}{James \surnamestart Leifer\surnameend} \&
  \bibinfo{author}{Robin \surnamestart Milner\surnameend}
  (\bibinfo{year}{2006}): \emph{\bibinfo{title}{Transition systems, link graphs
  and Petri nets}}.
\newblock {\sl \bibinfo{journal}{Mathematical Structures in Computer Science}}
  \bibinfo{volume}{16}(\bibinfo{number}{06}), pp. \bibinfo{pages}{989--1047},
  \doi{10.1017/S0960129506005664}.

\bibitemdeclare{article}{milner2006pure}
\bibitem{milner2006pure}
\bibinfo{author}{Robin \surnamestart Milner\surnameend} (\bibinfo{year}{2006}):
  \emph{\bibinfo{title}{Pure bigraphs: Structure and dynamics}}.
\newblock {\sl \bibinfo{journal}{Information and computation}}
  \bibinfo{volume}{204}(\bibinfo{number}{1}), pp. \bibinfo{pages}{60--122},
  \doi{10.1016/j.ic.2005.07.003}.

\bibitemdeclare{book}{Milner.Bigraphs.2009}
\bibitem{Milner.Bigraphs.2009}
\bibinfo{author}{Robin \surnamestart Milner\surnameend} (\bibinfo{year}{2009}):
  \emph{\bibinfo{title}{{The Space and Motion of Communicating Agents}}}.
\newblock \bibinfo{publisher}{Cambridge University Press},
  \doi{10.1017/CBO9780511626661}.

\bibitemdeclare{inproceedings}{perrone2012model}
\bibitem{perrone2012model}
\bibinfo{author}{Gian \surnamestart Perrone\surnameend},
  \bibinfo{author}{S{\o}ren \surnamestart Debois\surnameend} \&
  \bibinfo{author}{Thomas \surnamestart Hildebrandt\surnameend}
  (\bibinfo{year}{2012}): \emph{\bibinfo{title}{A model checker for bigraphs}}.
\newblock In: {\sl \bibinfo{booktitle}{Proc.\ Annual ACM Symposium on Applied
  Computing}}, \bibinfo{organization}{ACM}, pp. \bibinfo{pages}{1320--1325},
  \doi{10.1145/2245276.2231985}.

\bibitemdeclare{inproceedings}{delta2}
\bibitem{delta2}
\bibinfo{author}{Christopher \surnamestart Pietsch\surnameend},
  \bibinfo{author}{Timo \surnamestart Kehrer\surnameend}, \bibinfo{author}{Udo
  \surnamestart Kelter\surnameend}, \bibinfo{author}{Dennis \surnamestart
  Reuling\surnameend} \& \bibinfo{author}{Manuel \surnamestart
  Ohrndorf\surnameend} (\bibinfo{year}{2015}): \emph{\bibinfo{title}{SiPL - {A}
  Delta-Based Modeling Framework for Software Product Line Engineering}}.
\newblock In: {\sl \bibinfo{booktitle}{Proc.\ Intl.\ Conf.\ on Automated
  Software Engineering}}, \bibinfo{publisher}{{IEEE} Computer Society}, pp.
  \bibinfo{pages}{852--857}, \doi{10.1109/ASE.2015.106}.

\bibitemdeclare{article}{sevegnani2015bigraphs}
\bibitem{sevegnani2015bigraphs}
\bibinfo{author}{Michele \surnamestart Sevegnani\surnameend} \&
  \bibinfo{author}{Muffy \surnamestart Calder\surnameend}
  (\bibinfo{year}{2015}): \emph{\bibinfo{title}{{Bigraphs with Sharing}}}.
\newblock {\sl \bibinfo{journal}{Theor. Comput. Sci.}} \bibinfo{volume}{577},
  pp. \bibinfo{pages}{43--73}, \doi{10.1016/j.tcs.2015.02.011}.

\bibitemdeclare{book}{EMFBook}
\bibitem{EMFBook}
\bibinfo{author}{Dave \surnamestart Steinberg\surnameend},
  \bibinfo{author}{Frank \surnamestart Budinsky\surnameend},
  \bibinfo{author}{Ed~\surnamestart Merks\surnameend} \&
  \bibinfo{author}{Marcelo \surnamestart Paternostro\surnameend}
  (\bibinfo{year}{2008}): \emph{\bibinfo{title}{EMF: Eclipse Modeling
  Framework}}.
\newblock \bibinfo{publisher}{Pearson Education}.

\bibitemdeclare{inproceedings}{sescps}
\bibitem{sescps}
\bibinfo{author}{Christos \surnamestart Tsigkanos\surnameend},
  \bibinfo{author}{Timo \surnamestart Kehrer\surnameend},
  \bibinfo{author}{Carlo \surnamestart Ghezzi\surnameend},
  \bibinfo{author}{Liliana \surnamestart Pasquale\surnameend} \&
  \bibinfo{author}{Bashar \surnamestart Nuseibeh\surnameend}
  (\bibinfo{year}{2016}): \emph{\bibinfo{title}{Adding Static and Dynamic
  Semantics to Building Information Models}}.
\newblock In: {\sl \bibinfo{booktitle}{Proc.\ Intl.\ Workshop on Software
  Engineering for Smart Cyber-Physical Systems}}, \bibinfo{publisher}{ACM}, pp.
  \bibinfo{pages}{1--7}, \doi{10.1145/2897035.2897042}.

\bibitemdeclare{inproceedings}{tsigkanos2015ariadne}
\bibitem{tsigkanos2015ariadne}
\bibinfo{author}{Christos \surnamestart Tsigkanos\surnameend},
  \bibinfo{author}{Liliana \surnamestart Pasquale\surnameend},
  \bibinfo{author}{Carlo \surnamestart Ghezzi\surnameend} \&
  \bibinfo{author}{Bashar \surnamestart Nuseibeh\surnameend}
  (\bibinfo{year}{2015}): \emph{\bibinfo{title}{Ariadne: Topology Aware
  Adaptive Security for Cyber-Physical Systems}}.
\newblock In: {\sl \bibinfo{booktitle}{Intl.\ Conf.\ on Software Engineering}},
  \bibinfo{publisher}{{IEEE} Computer Society}, pp. \bibinfo{pages}{729--732},
  \doi{10.1109/ICSE.2015.234}.

\bibitemdeclare{article}{weiser1991computer}
\bibitem{weiser1991computer}
\bibinfo{author}{Mark \surnamestart Weiser\surnameend} (\bibinfo{year}{1991}):
  \emph{\bibinfo{title}{The computer for the 21st century}}.
\newblock {\sl \bibinfo{journal}{Scientific american}}
  \bibinfo{volume}{265}(\bibinfo{number}{3}), pp. \bibinfo{pages}{94--104},
  \doi{10.1145/329124.329126}.

\end{thebibliography}
